\documentclass[12pt]{article}%
\usepackage{amsfonts}
\usepackage{euscript,amsmath,amssymb,latexsym}%
\usepackage{amsmath}%
\usepackage{amssymb}%
\usepackage{graphicx}

\newcommand{\be}{\begin{equation}}
\newcommand{\ee}{\end{equation}}
\newcommand{\bea}{\begin{eqnarray}}
\newcommand{\eea}{\end{eqnarray}}

\newcommand{\Rh}{{\mathbb R}}

\newcommand{\const}{\mathrm{const}}
\renewcommand{\phi}{\varphi}

\newtheorem{theorem}{Theorem}[section]
\newtheorem{definition}{Definition}[section]
\newtheorem{example}{Example}
\newcommand*{\sign}{\mathrm{sign}}

\begin{document}

\title{\bf On Solutions to the Wave Equation on Non-globally Hyperbolic Manifold}
\vspace{1cm}
\author{
 O.V. Groshev\footnote{O.V. Groshev , M.V. Lomonosov Moscow State
University, GSP1, Leninskie Gory, 119991 Moscow,
Russia;email:\texttt{oleg.vic@gmail.com}}, N.A. Gusev\footnote{N.A.
Gusev, Moscow Institute of Physics and Technology, 9 Institutskii
per., 141700 Dolgoprudny, Moscow Region, Russia;
email:\texttt{nick.goussev@gmail.com}}, E.A. Kuryanovich\footnote{E.
A. Kuryanovich, email:\texttt{kurianovich@mail.ru}}, I.V.
Volovich\footnote{I.V. Volovich, Steklov Mathematical Institute, 8
ul. Gubkina, 119991 Moscow, Russia;
email:\texttt{volovich@mi.ras.ru}}}

\date {~}
\maketitle
\begin{abstract} We consider the Cauchy problem for the wave equation on a non-globally hyperbolic
manifold of the special form (Minkowski plane with a handle)
containing closed timelike curves (time machines). We prove that the
classical solution of the Cauchy problem exists and is unique if and
only if the initial data satisfy to some  set of additional
conditions.
\end{abstract}

\newpage

\section{Introduction}

 There is currently a quite developed theory of Cauchy problem for
 hyperbolic equations on globally hyperbolic
manifolds \cite{Adamar}--\cite{VS}. {\it Globally hyperbolic
manifold} is a spacetime oriented with respect to time (i.e., a pair
$(M,g)$, where $M$ is a smooth manifold and $g$ is the Lorentz
metric) if $M$ is diffeomorphic to $\mathbb{R}^1\times \Sigma$,
where $\Sigma$ is a Cauchy surface. This definition is equivalent to
Leray’s definition of global hyperbolicity
\cite{Leray,Hawking-Ellis}.

Hyperbolic equations on non-globally hyperbolic spacetimes have been
studied considerably less, although numerous examples of such
spacetimes are described by well-known solutions of gravitation
field equations; such are the solutions of G\"odel, Kerr, Gott and
many others  \cite{Hawking-Ellis}-\cite{GK}. These manifolds contain
closed timelike curves (time machine) and are non-globally
hyperbolic.

 Elementary examples of non-globally hyperbolic space–times are $\mathbb{S}^1_t\times \mathbb{R}^3_x$,
 with the Minkowski metric in which time argument passes a circle, and also anti-de Sitter space.
 There are several  papers where the hyperbolic equations on non-globally hyperbolic manifolds
 were discussed \cite{FMNEKTY}-\cite{Pol}.

Our purpose here is to study the wave equation on a manifold
containing closed time-like curves (CTC). We consider the Minkowski
plane with two slits whose edges are glued in a specific manner (
plane with a handle). In paper \cite{AVT} the Cauchy problem for the
wave equation on the Minkowski plane with handle was considered and
it was proved that there exists a solution, which is generally
discontinuous on the characteristics emerging from the conical
points.

In this work we establish the necessary and sufficient conditions on
the initial data for existence and uniqueness of the
\emph{classical} (i.e. smooth) solution to the Cauchy problem in the
half-plane $t\geqslant 0$ with exception of slits.

Our motivation is related with studying the possibility of creating
``wormholes'' and non-globally hyperbolic regions (mini time
machines) in collisions of the
 high-energy particles  \cite{TM}, also see \cite{MMT}.

Formation of CTC is related with the violation of the null energy
condition \cite{AreVol}.

Problems of boundary control for wave equation are considered in \cite{IM}.

Nonstandard boundary conditions for dynamic equations are discussed in \cite{KV,NN}.

We use in this work the following method to obtain the classical
solution: we divide upper half-plane into 7 regions $D_1,...,D_7$,
write out the general solutions of wave equation in each of these
regions and then we try to satisfy the gluing conditions and initial
data, solving a certain system of linear equations. The specific
conditions on the initial data \eqref{PreCond1}--\eqref{FinCond}
appear in this case.  Thus we obtain the classical solution to the
wave equation on the plane with handle (theorem 3.1). We also give
another method to solving problem using theory of disributions
(theorem 5.3). The results obtained by these two methods are
equivalent.

In this work we consider boundary value problems for the wave
equation on the Minkowski plane with the handle. It would be
interesting to establish relationship  with the known theory of the
boundary value problems for Laplace operator on the Riemann
surfaces, see for example, \cite{DNF,For}.

\section{Setting the problem}

We consider two vertical intervals $\gamma_1$ and $\gamma_2$ with
length $\ell>0$ in a half-plane
 $\mathbb{R}^2_+=\{(x,t)\in \mathbb{R}^2|t>0\}$:
 \be
 \label{gamma1} \gamma_1=\{(x,t)\in \mathbb{R}
_{+}^{2}|x=a_1,\,\,b_1<t<b_1+\ell\}\,, \ee \be \label{gamma2}
\gamma_2=\{(x,t)\in \mathbb{R} _{+}^{2}|x=a_2,\,\,b_2<t<b_2+\ell\}
\ee We suppose that \be a_2>a_1,\,\,\,b_2>b_1+\ell+a_2-a_1\,.
\label{equ:timelike} \ee

\begin{figure}[h]
$\,\,\,\,\,\,\,$$\,\,\,\,\,\,\,$$\,\,\,\,\,\,\,$
$\,\,\,\,\,\,\,$$\,\,\,\,\,\,\,$$\,\,\,\,\,\,\,$
$\,\,\,\,\,\,\,$$\,\,\,\,\,\,\,$$\,\,\,\,\,\,\,$
\includegraphics[height=2.5in]{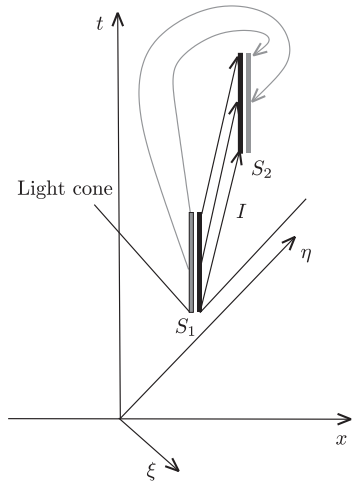}
\caption{\label{slits} \small Minkowski plane with two slits glued
in a specific way: the ``inner'' edges of the slits are glued one
with another, while the ``outer'' edgees of the slits are glued with
each other. The identifications of the ``outer'' and
  ``inner'' edge points are shown by arrows. There is also drawn a light cone emerging
out of the point $S_1$ with coordinates $(a_1,b_1)$. We assume that
the vector $I$ generating identifications is time-like.
  The point $S_2$ has coordinates $(a_2,b_2)$.
  }
\end{figure}

We assume that the edges of the intervals are glued as it is shown
$~$ in Fig.\ref{slits}. The resulting manifold has two conic points
-- the ends of the intervals.

Every continuous field  on this manifold will satisfy certain gluing
conditions on the slits $\gamma_1$ and $\gamma_2$. Conversely, if
the field is continuous in domain $\Omega = \Rh^2_+ \setminus
\bar\gamma_1 \cup \bar\gamma_2$ and satisfies those gluing
conditions then it is continuous on the manifold.

Consider the wave equation on that manifold for the function
$u=u(x,t)$
\be \label{FIRST}
  u_{tt} - u_{xx} = 0, ~ (x,t) \in \Omega
\ee with initial conditions \bea\label{Cau}
 u(x,0) &=& \phi(x)\,,\\
 u_t(x,0) &=& \psi(x)\,,\label{MIDDLE}
\eea where $\phi\in C^2(\Rh)$, $\psi\in C^1(\Rh)$. Let us set the
following gluing conditions: \bea \label{surg}
 u(a_1-0, t) &=& u(a_2+0,t+b_2-b_1 )\,,\\
 u(a_1+0, t) &=& u(a_2-0,t+b_2-b_1)\,,\\
 u_x(a_1-0, t) &=& u_x(a_2+0,t+b_2-b_1)\,,\\
\label{LAST}
 u_x(a_1+0, t) &=& u_x(a_2-0,t+b_2-b_1),
\eea where $b_1 < t < b_1 + \ell$ and the indicated limits exist. We
will show below that no extra conditions needed.

Let us define the classical solution:

{\bf Definition 2.1.}
 A function $u \in C^2(\Omega) \cap C^1 (\Omega\cup\{t=0\})$ is called the classical
solution to the problem \eqref{FIRST}--\eqref{LAST} if it satisfies
conditions \eqref{FIRST}--\eqref{LAST}, provided the indicated left-
and right-hand side limits exist.
$$\,$$
\begin{figure}[h]
\center{\includegraphics[height=2.1in]{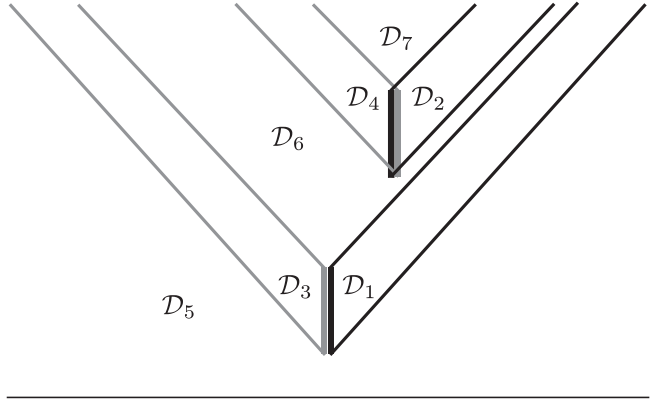}}
\caption{\label{domains} \small Domains $D_1$,...,$D_7$
  }
\end{figure}
Using characteristic half-lines emerging out of the ends of the intervals
$\bar\gamma_1 , \bar\gamma_2$, we divide the upper half-plane $\Rh^2_+$
into 7 simply connected domains $D_1$,...,$D_7$ (see Fig.2):
\begin{eqnarray*} \label{D_i}
D_{i}:&x>a_{i},a_{i}-b_{i}-l<x-t<a_{i}-b_{i},\\
 D_{i+2}:& x<a_{i},a_{i}+b_{i}<x+t<a_{i}+b_{i}+l,\; i=1,2,\\
 D_{5}:& 0<t<|x-a_{1}|+b_{1},\\
 D_{6}:& |x-a_{1}|+b_{1}+l<t<|x-a_{2}|+b_{2},\\
 D_7:& t>|x-a_{2}|+b_{2}+l.\end{eqnarray*}

It will be shown that if the initial data meet certain conditions,
 then  the classical solution to the problem
\eqref{FIRST}--\eqref{LAST} exists, is unique, and is of the form\be
 \label{General}
 u(x,t)=f(x-t+A_i)+g(x+t+B_i)+C_i
 \ee
 in each domain $D_i,~i=1,...,7$, where  $A_i,B_i,C_i$ are constants
 and $f$, $g$ are functions defined by d'Alembert's fomulae.

\section{Theorem of existence of classical solution}



 Hereafter we will use notations:
$$
 a = a_2 - a_1,\quad b = b_2 - b_1,
$$
and
$$
 c_i = a_i - b_i,\quad d_i = a_i + b_i,
$$
where $i$ can be 1, 2 or empty (in particular, $c=a_2 - a_1-b_2 +
b_1$).

We will prove that the existence of a classical solution is
equivalent to fulfilling the conditions \be
 \label{PreCond1}
  \phi(c_2-\ell) - \phi(c_1-\ell) + \int^{c_1-\ell}_{c_2-\ell} \psi(s) \,ds=
  \phi(d_1) - \phi(d_2) - \int^{d_2}_{d_1} \psi(s) \,ds,
\ee

\be
 \label{PreCond2}
  \phi(d_2+\ell) - \phi(d_1+\ell) + \int^{d_2+\ell}_{d_1+\ell} \psi(s) \,ds=
  \phi(c_1) - \phi(c_2) - \int^{c_1}_{c_2} \psi(s) \,ds.
\ee
and conditions of smoothness on characteristics
\begin{eqnarray}
\label{PreCond3}
 \phi^{(i)}(c_1) - \psi^{(i-1)}(c_1) &=& \phi^{(i)}(c_2) - \psi^{(i-1)}(c_2)\\
 \phi^{(i)}(d_1) + \psi^{(i-1)}(d_1) &=& \phi^{(i)}(d_2) + \psi^{(i-1)}(d_2)\\
 \phi^{(i)}(c_1-\ell) - \psi^{(i-1)}(c_1-\ell) &=& \phi^{(i)}(c_2-\ell) - \psi^{(i-1)}(c_2-\ell)\\
\label{FinCond}
 \phi^{(i)}(d_1+\ell) + \psi^{(i-1)}(d_1+\ell) &=& \phi^{(i)}(d_2+\ell) + \psi^{(i-1)}(d_2+\ell),\; i=1,2.
\end{eqnarray}
Namely, the following theorem holds.

{\bf Theorem 3.1.} {\it The classical solution to the problem
\eqref{FIRST}--\eqref{LAST} exists if and only if the conditions
\eqref{PreCond1}--\eqref{FinCond} for $\phi, \psi$ hold. Given this,
the classical solution is unique and is given by the formula
\begin{equation}\label{equ:classic}
u(x,t) = u_i(x,t) \quad \mbox{if} \quad
(x,t)\in D_i,\; i=1,\dots,7,
\end{equation}
 where
\begin{eqnarray}
\label{Equa1}
 u_1(x,t) &=& f(\eta+c) + g(\xi) + f(c_1) - f(c_2)\\
 u_2(x,t) &=& f(\eta-c) + g(\xi) + g(d_1) - g(d_2)\\
 u_3(x,t) &=& f(\eta) + g(\xi+d) + g(d_1) - g(d_2)\\
 u_4(x,t) &=& f(\eta) + g(\xi-d) + f(c_1) - f(c_2)\\
 u_5(x,t) &=& f(\eta) + g(\xi)\\
 u_6(x,t) &=& f(\eta) + g(\xi) + g(d_1) - g(d_2) + f(c_1) - f(c_2)\\
 \label{Equa7}
 u_7(x,t) &=& f(\eta) + g(\xi);
\end{eqnarray}
 here $\xi = x+t,\; \eta = x-t$,
\begin{equation}
 f(x) = \frac12\left[ \phi(x) - \int_{x_0}^x \psi(s) \,ds \right]
\end{equation}
and
\begin{equation}
 g(x) = \frac12\left[ \phi(x) + \int_{x_0}^x \psi(s) \,ds \right].
\end{equation}
}

 {\it Proof.}  An arbitrary solution to equation \eqref{FIRST} in domain $D_i, i=1,...,7$ is given by
$$
  u(x,t) = f_i(x-t) + g_i(x+t), \qquad\mbox{when }(x,t) \in
  D_i,~i=1,...,7,
$$
where $f_i(x-t),~g_i(x+t) \in C^2(D_i),~ i=1,...,7$. We will show
that conditions \eqref{Cau}--\eqref{LAST} impose quite strong
restrictions to $f_i$ and $g_i$.

Functions $f_5$ and $g_5$ are calculated directly from $\phi, \psi$
via the d'Alembert formulae:
\begin{equation}
 f_5(x) = f(x),\quad g_5(x) = g(x).
\end{equation}

From now on we will evaluate $f_i, g_i$ through $f,
g$ in a manner to make the solution $u$ twice continuously differentiable on  $\Omega$,
including eight characteristics $\Gamma_{ij} = \overline D_i
\cap \overline D_j \cap \Omega$; here $i, j$ take up such values from $1,\dots,7$
that $\Gamma_{ij}$ is open half-line.
Let us write continuity conditions on $\Gamma_{51}$, i.~e.
\begin{equation}
\label{cont}
 u_1 = u_5,\quad (x,t) \in \Gamma_{51}.
\end{equation}
Analytically half-line $\Gamma_{51}$ is given by $\{ x-a_1 = t-b_1 > 0
\}$. Thus we can write \eqref{cont} as
$$
 f_1(a_1-b_1) + g_1(2t+a_1-b_1) = f_5(a_1-b_1) + g_5(2t+a_1-b_1), \quad t > b_1.
$$
Using our notations, we evaluate $g_1$:
$$
 g_1(2t+c_1) = g_5(2t+c_1) + G_{51}, \quad t > b_1,
$$
where
$$
 G_{51} = f_5(c_1) - f_1(c_1).
$$

Therefore, we have defined function $g_1(\xi)$ when $\xi > a_1+b_1$;
thus it is also defined when $(x,t) \in D_1$; in addition, $g_1(\xi)$ equals
$g(\xi)$ up to constant.

Similarly, using continuity conditions on $\Gamma_{16},\;
\Gamma_{62},\; \Gamma_{27}$ we get functions $g_6,\; g_2,\;g_7$ defined
when $\xi$ is greater than $d_1+\ell,\; d_2,\; d_2+\ell$ respectively and equal $g(\xi)$
up to constant.

In a similar way, it is easy to show that functions $f_3,\; f_6,\;
f_4,\; f_7$ of $\eta$ are defined when $\eta > c_1,\; c_1-\ell,\;
c_2,\; c_2-\ell$ and are equal to $f(\eta)$ up to constants.

 {\bfseries Gluing conditions.} Now we apply the gluing conditions for functions \eqref{surg}:
$$
 u_1(a_1, t) = u_4( a_2,t+b),
$$
i.~e.
\begin{equation}
 \label{func}
  f_1(a_1-t) + g_1(a_1+t) = f_4(a_2-(t+b)) + g_4(a_2+t+b).
\end{equation}
And gluing conditions for derivatives are
\begin{equation}
\label{deri}
 f'_1(a_1-t) + g'_1(a_1+t) = f'_4(a_2-(t+b)) + g'_4(a_2+t+b).
\end{equation}

Let us differentiate \eqref{func} on $t$ and add it to \eqref{deri}.
We obtain
$$
 g'_1(a_1+t) = g'_4(a_2+t+b).
$$
Thus,
$$
 g_4(\xi) = g_1(\xi-a-b) + \const.
$$
Let us note that as this equation holds for
$$
 a_2 + b_2 < \xi < a_2 + b_2 + \ell,
$$
it defines $g_4(x+t)$ for $(x,t) \in D_4$.

From \eqref{func} we obtain
$$
 f_1(a_1-t) = f_4(a_2-t-b) + \const.
$$
Therefore, the function
$$
 f_1(\eta) = f_4(\eta + a-b) + \const,
$$
is defined for all $\eta = x-t$ when $(x,t) \in D_1$.

Finally, we have
$$
 g_4(\xi) = g_1(\xi-d)+\const
$$
and
$$
 f_1(\eta) = f_4(\eta + c)+\const.
$$

Similarly, using the gluing conditions for $u_2$ and $u_3$, we have
$$
 g_3(\xi) = g_2(\xi+d)+\const
$$
and
$$
 f_2(\eta) = f_3(\eta-c)+\const.
$$

{\bfseries Evaluating constants.} We have obtained solution in the form
\begin{eqnarray}
\label{WOConst1}
 u_1(x,t) &=& f(\eta+c) + g(\xi) + U_1,\\
 u_2(x,t) &=& f(\eta-c) + g(\xi) + U_2,\\
 u_3(x,t) &=& f(\eta) + g(\xi+d) + U_3,\\
 u_4(x,t) &=& f(\eta) + g(\xi-d) + U_4,\\
 u_5(x,t) &=& f(\eta) + g(\xi),\\
 u_6(x,t) &=& f(\eta) + g(\xi) + U_6,\\
\label{WOConst7}
 u_7(x,t) &=& f(\eta) + g(\xi) + U_7.
\end{eqnarray}
Now we have to find the constants $U_i$.

It follows from \eqref{func} that
$$
 U_1 = U_4;
$$
similarly,
$$
 U_2 = U_3.
$$

Now we will find $U_1$ and $U_2$, by employing the continuity
conditions for solution on the half-lines $\Gamma_{51}$ and
$\Gamma_{53}$ respectively. We have $\eta = c_1$ on $\Gamma_{51}$,
so we can write
$$
 u_1 = u_5
$$
as
$$
 f(c_1 + c) + g(\xi) + U_1 = f(c_1) + g(\xi).
$$
Recalling $c_1+c=c_2$, we have
$$
 U_1 = f(c_1) - f(c_2).
$$
In a similar manner we get
$$
 U_2 = g(d_1) - g(d_2).
$$
Now we consider the half-lines $\Gamma_{16}$ and $\Gamma_{36}$.
Continuity condition on $\Gamma_{16}$ is written as
$$
 f(c_1-\ell+c) + g(\xi) + U_1 = f(c_1-\ell) + g(\xi) + U_6,
$$
wherefrom
$$
 U_6 = f(c_2-\ell) - f(c_1-\ell) + f(c_1) - f(c_2).
$$
Similarly, the continuity on $\Gamma_{36}$ is written as
$$
 f(\eta) + g(d_1+\ell+d) + U_2 = f(\eta) + g(d_1+\ell) + U_6,
$$
wherefrom, bearing in mind $d_1+d=d_2$, we get
$$
 U_6 = g(d_2+\ell) - g(d_1+\ell) + g(d_1) - g(d_2).
$$

We have obtained the condition for the functions $f$, $g$:
\begin{equation}
\label{C6}
 f(c_2-\ell) - f(c_1-\ell) + f(c_1) - f(c_2) = g(d_2+\ell) - g(d_1+\ell) + g(d_1) - g(d_2).
\end{equation}
As we will notice, we need two conditions for the continuous
solution; the obtained condition will necessarily follow from those
two.

So, let us consider the half-lines $\Gamma_{62}$ and $\Gamma_{64}$.
We have $\eta = c_2$ on $\Gamma_{62}$. Let us insert it into
$$
 u_6 = u_2.
$$
We get
$$
 f(c_2) + U_6 = f(c_2-c) + U_2.
$$
Inserting the found constants, we get
$$
 f(c_2) + f(c_2-\ell) - f(c_1-\ell) + f(c_1) - f(c_2) = f(c_1) + g(d_1) - g(d_2).
$$
Thus we have found the first condition:
\begin{equation}
\label{C1}
 f(c_2-\ell) - f(c_1-\ell) = g(d_1) - g(d_2).
\end{equation}
If we express $f$,  $g$ through $\phi, \psi$, we will have exactly
\eqref{PreCond1}.

Consider $\Gamma_{64}$. We have $\xi = d_2$ on it; computing
similarly, we obtain the second condition
\begin{equation}
\label{C2}
 f(c_1) - f(c_2) = g(d_2+\ell) - g(d_1+\ell).
\end{equation}
Easy to see that if we add \eqref{C1} to \eqref{C2} we will obtain
precisely the condition \eqref{C6}.

We are left to find the last constant $U_7$. We consider
conditions on $\Gamma_{27}$: let us insert $\eta = c_2 - \ell$ into
$$
 u_2|_{\Gamma_{27}}(\eta) = u_7|_{\Gamma_{27}}(\eta).
$$
We obtain
$$
 f(c_1-\ell) + U_2 = f(c_2-\ell) + U_7.
$$
Recalling \eqref{C1}, we get
$$
 U_7 = 0.
$$
One can easily check that the continuity condition on $\Gamma_{47}$
also yields zero $U_7$.

Hence, inserting obtained $U_i$ into
\eqref{WOConst1}--\eqref{WOConst7}, we get the solution $u$ given by
\eqref{Equa1}--\eqref{Equa7}.

{\bfseries Differentiability conditions.} We will find the
conditions for differentiability of the solutions on the half-lines
$\Gamma_{ij}$. The partial derivatives along half-lines
$\Gamma_{ij}$ exist, as it follows directly from the formulae
\eqref{Equa1}--\eqref{Equa7}. Let us write the conditions for
continuity of partial derivatives of solution along normals to
corresponding half-lines.
\begin{eqnarray}
 \label{CondFrst}
 f^{(i)}(c_1) &=& f^{(i)}(c_2)\\
 g^{(i)}(d_1) &=& g^{(i)}(d_2)\\
 f^{(i)}(c_1-\ell) &=& f^{(i)}(c_2-\ell)\\
 \label{CondLast}
 g^{(i)}(d_1+\ell) &=& g^{(i)}(d_2+\ell), \; i = 1,2.
\end{eqnarray}
These conditions are equivalent to \eqref{PreCond3}--\eqref{FinCond}.

Theorem 3.1 is proved. $\blacksquare $

\subsection{Example}

We will discuss an example when all conditions of the theorem are
satisfied, and thus, the classical solution exists. We will look for
the solution of the right-mode form:
$$
 u = f(x-t).
$$

From $g \equiv 0$ it follows that we should pick such initial
conditions:
$$
 \psi = -\phi'.
$$
Then $f=\phi$. We choose as $\phi$ bump function with
support in $[c_1-\ell, c_1]$:
$$
 \phi (x)= \begin{cases}\exp\left( -\frac{\ell^2}{ \ell^2 -
 4(x-c_1+\ell/2)^2} \right), & x\in (c_1-\ell, c_1),\\
                            0, & x \notin {(c_1-\ell, c_1)} \end{cases}
$$

 Conditions \eqref{PreCond1}--\eqref{FinCond} are fulfilled. The solution is
right-travelling wave, coming into the lower slit and leaving out of the upper one.

\section{Discontinuity jumps at slits}

In the next section we will study the problem
\eqref{FIRST}--\eqref{LAST} by means of theory of distributions. We
will generalize the method of analysis of the Cauchy problem from
\cite{VS} to our case of plane with slits. Our method can be of
interest in the analysis of generalized solutions of the problem
concerned. Here we shall confine ourselves to study some properties
of classical solutions of problem \eqref{FIRST}--\eqref{LAST} in the
``strengthened'' setting.

We will use the following notations for the ``one-sided'' limits and
discontinuity jumps of functions:
\begin{equation}\label{equ:def:limits}
\begin{gathered}
(x,t)\to (A-0,B) \quad \Leftrightarrow \quad (x,t)\to (A,B) \; | \; x<A \\
(x,t)\to (A+0,B) \quad \Leftrightarrow \quad (x,t)\to (A,B) \; | \; x>A \\
[F(x,t)]_{x=A}\equiv[F]_{x=A}(t)= \lim_{(x,\tau)\to (A+0, t)}
F(x,\tau) - \lim_{(x,\tau)\to (A-0, t)} F(x,\tau).
\end{gathered}
\end{equation}

For convenience we shall introduce the following class $\mathcal{K}$
of functions:

\begin{definition}
A function $u(x,t)$ belongs to the class $\mathcal{K}$ if $u(x,t)\in
C^2(\Omega) \cap C^1 (\Omega\cup\{t=0\})$ and there exist the
following limits:
 \begin{gather*}
  \lim_{\textstyle(x,\tau)\to (a_i \pm 0, b_i+ t)} Du(x,\tau),
 \end{gather*}
where $i=1,2$, $Du = \{u, u_x, u_t\}$, $0\leqslant t \leqslant \ell$.
\end{definition}

\begin{definition}[``strengthened'' setting of problem \eqref{FIRST}--\eqref{LAST}]
\label{problem:2} The solution $u(x,t)$ of the problem
\eqref{FIRST}--\eqref{MIDDLE} is called \emph{strengthened classical
solution of the problem \eqref{FIRST}--\eqref{LAST}} if $u(x,t)\in
\mathcal{K}$ and the following conditions are satisfied:
\begin{gather}
\lim_{(x,\tau)\to (a_1-0, b_1 + t)} u(x,\tau)=\lim_{(x,\tau)\to
(a_2+0, b_2 + t)} u(x,\tau),
\label{equ:cond1:1}\\
\lim_{(x,\tau)\to (a_1+0, b_1 + t)} u(x,\tau)=\lim_{(x,\tau)\to
(a_2-0, b_2 + t)} u(x,\tau),
\label{equ:cond2:1}\\
\lim_{(x,\tau)\to (a_1-0, b_1 + t)} u_x(x,\tau)=\lim_{(x,\tau)\to
(a_2+0, b_2 + t)} u_x(x,\tau),
\label{equ:cond1:2}\\
\lim_{(x,\tau)\to (a_1+0, b_1 + t)} u_x(x,\tau)=\lim_{(x,\tau)\to
(a_2-0, b_2 + t)} u_x(x,\tau), \label{equ:cond2:2}
\end{gather}
where $t\in[0,\ell]$.
\end{definition}

It is not difficult to see that the conditions
\eqref{surg}--\eqref{LAST} are weaker than the conditions
\eqref{equ:cond1:1}--\eqref{equ:cond2:2}.

Let us formulate the main properties of functions from class
$\mathcal{K}$ which comply with the conditions
\eqref{equ:cond1:1}--\eqref{equ:cond2:2}:
\begin{theorem}\label{th:K}
 Let $u(x,t)\in\mathcal{K}$ satisfies the conditions
 \eqref{equ:cond1:1}--\eqref{equ:cond2:2}.
 Let $\nu(t)$ and $\omega(t)$ denote the discontinuity jumps of function $u(x,t)$ and its derivative
 $u_x(x,t)$ at the upper slit $\gamma_2$ respectively:
\begin{gather*}
 \nu(t) = [u]_{x=a_2}(b_2 + t), \quad
 \omega(t) = [u_x]_{x=a_2}(b_2 + t).
\end{gather*}
 Then one has:
\begin{enumerate}
 \item \label{th:K:0} $u(x,t)\in L_{1,loc}(\mathbb R^2_{t\geqslant 0})$.
 \item \label{th:K:1} $\omega(t)\in C(\mathbb R)$, $\nu(t)\in C^1(\mathbb R)$, and
 for the discontinuity jumps at the lower slit $\gamma_1$ we have
 \begin{gather*}
 [u]_{x=a_1}(b_1+t) = -\nu(t), \\
 [u_x]_{x=a_1}(b_1+t) = -\omega(t),
 \end{gather*}
 moreover for $t\notin[0,\ell]$ we have $\nu(t)=\omega(t)=0$.

 \item Time derivatives satisfy the following gluing conditions:
\begin{equation}\label{equ:cond12:3}
\begin{gathered}
\lim_{(x,\tau)\to (a_1-0, b_1 + t)} u_\tau(x,\tau)=\lim_{(x,\tau)\to (a_2+0, b_2 + t)} u_\tau(x,\tau),\\
\lim_{(x,\tau)\to (a_1+0, b_1 + t)} u_\tau(x,\tau)=\lim_{(x,\tau)\to
(a_2-0, b_2 + t)} u_\tau(x,\tau),
\end{gathered}\end{equation}
where $t\in[0,\ell]$.
\end{enumerate}
\end{theorem}

Note that the conditions \eqref{equ:cond12:3}, in contrast to the
conditions \eqref{equ:cond1:1}--\eqref{equ:cond2:2}, are imposed on
time derivatives instead of space derivatives. Hence, in the
``strengthened'' setting of the problem there is no need in
additional gluing conditions for the solution $u(x,t)$ at the slits
$\gamma_1$ and $\gamma_2$.

\section{Classical and generalized solutions}
In this section we will derive the equation which will be satisfied
by every strengthened classical solution of the problem
\eqref{FIRST}--\eqref{LAST} in sense of distributions $\mathcal
D'(\mathbb R^2)$. We will use the following notation for the
d'Alembert operator: $\square\equiv\partial^2_t -
\partial^2_x.$ Also let $\mathcal D'_+(\mathbb R^2)$ denote
the set of distributions from $\mathcal D'(\mathbb R^2)$ which equal to 0
for $t<0$.

\begin{theorem}
Let $u(x,t)$ be a strengthened classical solution of the problem
\eqref{FIRST}--\eqref{LAST}. Then the function
$$\widetilde u (x,t)=\begin{cases}
 u(x,t), & t\geqslant 0, \\
 0, & t<0.
\end{cases}
$$ satisfies the following equation in the sense of $D'(\mathbb R^2)$:

\be\label{weF} \square \widetilde u(x,t) = F(x,t), \ee where
\be\label{defF} F(x,t)= \phi (x) \cdot \delta'(t) + \psi(x) \cdot
\delta(t) -
 [u]_{x=a_1}\cdot \delta'(x-a_1) - [u_x]_{x=a_1}\cdot \delta(x-a_1) \ee
 $$-[u]_{x=a_2}\cdot \delta'(x-a_2) - [u_x]_{x=a_2}\cdot \delta(x-a_2)
$$
\end{theorem}

The proof is similar to the derivation of the generalized Cauchy problem setting
given in \cite{VS}. It relies on the fact that
 $\widetilde u(x,t)\in L_{1,loc}(\mathbb R^2_{t\geqslant 0})$, which follows
 from Theorem \ref{th:K}.

Recall the following formula \cite{VS}: \be\label{Lapl}\triangle
f=\{\triangle f\}+\left[\frac{\partial f}{\partial
n}\right]_S\delta_S+\frac{\partial}{\partial n}([f]_S\delta_S),\ee
in sense of $D'(\mathbb{R}^n)$. Here $\triangle$ denotes the Laplace
operator in $\mathbb{R}^n$, function $f\in C^2 (\bar{G})\bigcap C^2
(\bar{G}_1)$, domain $G$ in $\mathbb{R}^n$ has partially smooth
boundary $S$, $G_1=\mathbb{R}^n \backslash \bar{G}$, $\{\triangle
f\}$ denotes the action of the classical Laplace operator and
$[f]_S$ denotes the discontinuity jump of $f$ at the surface $S$. We
have obtained the analog of this formula for the d'Alembert operator
on the plane with the slits.

Next, by virtue of theorem \ref{th:K}
\begin{equation}\label{equ:cond:delta}
\begin{gathered}{}
[u]_{x=a_1}(b_1+t) = -[u]_{x=a_2}(b_2 + t) = -\nu(t)\in C^1(\mathbb R);
\quad \nu(t)=0, ~ t\notin[0,\ell]\\
[u_x]_{x=a_1}(b_1+t) = -[u_x]_{x=a_2}(b_2 + t) = -\omega(t)\in
C(\mathbb R);
 \quad \omega(t)=0, ~ t\notin[0,\ell].
\end{gathered}
\end{equation}

Hence, in the ``strengthened'' setting problem \eqref{FIRST}--\eqref{LAST}
is equivalent to the following problem:

{\itshape Find functions $\omega(t)\in C(\mathbb R)$ and $\nu(t)\in
C^1(\mathbb R)$, equal to 0 for $t\notin[0,\ell]$, such that the
generalized solution in $D'(\mathbb R^2)$ of the equation
\begin{equation}\label{equ:general_wave}
\square  u(x,t) = F(x,t),
\end{equation}
\begin{equation}\label{equ:general_wave2}
\begin{gathered}
F(x,t) = \phi(x) \cdot \delta'(t) + \psi(x) \cdot \delta(t) +
 \nu(t-b_1)\cdot \delta'(x-a_1) + \omega(t-b_1)\cdot \delta(x-a_1) - \\
-\nu(t-b_2)\cdot \delta'(x-a_2) - \omega(t-b_2)\cdot \delta(x-a_2),
\end{gathered}
\end{equation}
belongs to class $\mathcal{K}$ and satisfies conditions
\eqref{equ:cond1:1} and \eqref{equ:cond2:2}. }

Note that the conditions \eqref{equ:cond1:2} and \eqref{equ:cond2:1}
will be satisfied automatically by virtue of conditions
\eqref{equ:cond:delta}.

So, the problem of existence and uniqueness of solution $u(x,t)$
has converted to the problem of existence and uniqueness of the discontinuity
jumps $\omega(t)$ and $\nu(t)$ satisfying specific conditions.
To obtain these conditions we will first find the general solution of equation
\eqref{equ:general_wave}.

\subsection{Solution of equation \eqref{equ:general_wave}}

As is known \cite{VS}, the solution of the generalized Cauchy problem
for equation \eqref{equ:general_wave} exists, is unique and is given
by a convolution of the fundamental solution $\mathcal E_1 $
with the right hand side $F$ defined in \eqref{equ:general_wave2}:
 \be\label{fundsol}u(x,t)=\mathcal E_1 *
F(x,t).\ee Here $$\mathcal E_1(x,t) = \frac{1}{2}\theta(t-|x|)$$
is the fundamental solution of operator $\square$, where $\theta (t)$ denotes
Heaviside step function; $\theta (t)=1$ for $t>0$  and $\theta (t)=0$ for
$t<0$.

Let us write out an explicit formula for the convolution
\eqref{fundsol}. For this purpose we use the following formulae:
\be\label{relat} \mathcal E_1 * \phi(x) \delta'(t)=\frac{1}{2}[\phi
(x+t)+\phi (x-t)], \ee
$$\mathcal E_1 *
\psi(x) \delta(t)=\frac{1}{2}\int_{x-t}^{x+t}\psi(s)ds,
$$
$$
\mathcal E_1 * \omega(t) \delta(x) =
\frac{1}{2}\theta(t-|x|)\int_0^{t-|x|} \omega(\tau)\,d\tau,
$$
$$
\mathcal E_1 * \nu(t) \delta'(x) = \frac{\partial}{\partial
x}[\mathcal E_1 * \nu(t) \delta(x)] =
-\theta(t-|x|)\frac{\mathrm{sign\,}{x}}{2} \nu\left(t-|x|\right).
$$
Therefore denoting \begin{equation}\label{equ:solUU}U(x,t)=
\frac{1}{2}\theta(t-|x|)\int_0^{t-|x|} \omega(\tau)\,d\tau
-\theta(t-|x|)\frac{\mathrm{sign\,}{x}}{2}
\nu\left(t-|x|\right),\end{equation} we obtain the solution of equation
\eqref{equ:general_wave} in the following form:
\begin{equation}\label{equ:solution2}
u(x,t)= u^D(x,t) + U(x-a_1,t-b_1) - U(x-a_2,t-b_2).
\end{equation}
Here $u^D$ denotes the solution of classical Cauchy problem for wave equation
defined by d'Alembert's fomula:
\begin{equation}
u^D(x,t)= \frac{1}{2}[\phi (x+t)+\phi
(x-t)]+\frac{1}{2}\int_{x-t}^{x+t}\psi(s)ds=f(x-t) + g(x+t).
\end{equation}

\subsection{Gluing conditions}

Let us now define the functions $\nu (t)$ and $\omega (t)$ using
the gluing conditions. Conditions at the slits \eqref{equ:cond1:1} и
\eqref{equ:cond2:2} take the form

\be\label{schiv1}
 u^D(a_1, b_1+t)+U(-0,t) - U(a_1-a_2-0,b_1-b_2+t) =\ee
 $$= u^D(a_2, b_2+t)+U(a_2-a_1,b_2-b_1+t) - U(+0, t),$$

 \be\label{schiv2}
 u^D_x(a_1, b_1+t)+U_x(-0,t)  = u^D_x(a_2, b_2+t)-U_x(+0, t),\ee

where $0 < t < \ell$.

Note that from \eqref{equ:solUU} follows that
$$
U(a_1-a_2-0,b_1-b_2+t)=0 ,
$$
because $-b_2+b_1+a_2-a_1+\ell <0$ (see \eqref{equ:timelike}) and that
$U(a_2-a_1,b_2-b_1+t)=const$ for $0 < t < \ell$, precisely:
$$
U(a_2-a_1,b_2-b_1+t)=\frac{1}{2}\int_0^{\ell} \omega(\tau)\,d\tau.
$$
We also have (``one-sided'' limits are meant is sense of \eqref{equ:def:limits}):
\begin{equation}\label{UUx}
 U(\pm 0,t)=\frac{1}{2}\int_0^t \omega (\tau ) d\tau\mp\frac{1}{2}\nu (t),
\end{equation}
$$
U_x(\pm 0,t)=\mp\frac{1}{2} \omega (t)+\frac{1}{2}\nu' (t).
$$

Therefore gluing conditions \eqref{schiv1} and \eqref{schiv2} take the form
\begin{equation}\label{equ:omega,nu-problem}
 \int_0^t\omega(\tau)d\tau=  \frac{1}{2}\int_0^{\ell}\omega(\tau)d\tau+ D_1(t),
 \end{equation}
 \begin{equation}\label{equ:omega,nu-problem2}
 \nu'(t) = D_2(t),
\end{equation}
where $$D_1(t) = u^D(a_2, b_2+t) - u^D(a_1, b_1+t),$$ $$ D_2(t) =
u^D_x(a_2, b_2+t) - u^D_x(a_1, b_1+t).$$

Problems \eqref{equ:omega,nu-problem},
\eqref{equ:omega,nu-problem2} have unique solutions respectively
\begin{equation}\label{equ:omega,nu}
 \omega(t)=D_1'(t),
 \end{equation}
 \begin{equation}\label{equ:omega,nu2}
 \nu(t) = \int_0^t D_2(\tau)\, d\tau.
\end{equation}
These solutions are sufficiently smooth
(recall that $\omega(t)\in C^1(\mathbb R)$,
$\nu(t)\in C^2(\mathbb R)$ and also $\omega(t)=\nu(t)=0$ for $t\notin [0,\ell]$)
when and only when the following conditions are satisfied:
\begin{equation}\label{equ:solvability}
  D_1'(0) = D_1'(\ell) = 0, \quad D_1(0) + D_1(\ell) = 0,
  \quad D_1''(0) = D_1''(\ell) = 0,
  \end{equation}
  \begin{equation}\label{equ:solvability2}
  \int_0^\ell D_2(\tau)\, d\tau = 0, \quad D_2(0) = D_2(\ell) = 0,
   \quad D_2'(0) = D_2'(\ell) = 0,
\end{equation}
These conditions are derived by direct substitution $t=0,\ell$ into
\eqref{equ:omega,nu-problem}, \eqref{equ:omega,nu-problem2}.

Therefore we have obtained the following result:
\begin{theorem}
There exists a unique strengthened classical solution of the problem
\eqref{FIRST}--\eqref{LAST} if and only if the conditions
\eqref{equ:solvability}, \eqref{equ:solvability2} to the initial
data  are satisfied. This solution is given by
\begin{equation}\label{equ:solution}
u(x,t)= u^D(x,t) + U(x-a_1,t-b_1) - U(x-a_2,t-b_2),
\end{equation}
where
\begin{gather*}
u^D(x,t)= \frac{1}{2}[\phi (x+t)+\phi
(x-t)]+\frac{1}{2}\int_{x-t}^{x+t}\psi(s)ds ,\\
U(x,t)=\frac{1}{2}\theta(t-|x|)\int_0^{t-|x|} \omega(\tau)\,d\tau
-\theta(t-|x|)\frac{\mathrm{sign\,}{x}}{2} \nu\left(t-|x|\right),\\
\omega(t) = \theta(t)\theta(\ell-t)\cdot\left(u^D_t(a_2, b_2+t) -
u^D_t(a_1, b_1+t)\right),\\
\nu(t)=\theta(t)\theta(\ell-t)\cdot\int_0^t \left( u^D_x(a_2,
b_2+\tau) - u^D_x(a_1, b_1+\tau)\right) d\tau  .
\end{gather*}
\end{theorem}

One can show that
\begin{itemize}
 \item the conditions \eqref{equ:solvability} are equivalent to the conditions
 \eqref{PreCond1}--\eqref{FinCond};
 \item the strengthened classical solution $u(x,t)$ given by
 \eqref{equ:solution} is identical to the classical solution given by \eqref{equ:classic}.
\end{itemize}

Note that if we drop the gluing conditions \eqref{equ:cond1:2} and \eqref{equ:cond2:2} for the derivative
$u_x$ then the solution of the problem concerned will not be unique. Indeed,
in this case we can substitute arbitrary $\omega(t)$ (such that $\omega(t)\in C^1(\mathbb R)$ and also
$\omega(t)=0$ for $t\notin [0,\ell]$) into \eqref{equ:solution}.

Let us present an example (belonging to T. Ishiwatari) of nontrivial
solution $u^D$ and parameters $a_i$, $b_i$, for which conditions
\eqref{equ:solvability} are satisfied:

\begin{example}
 Let $a_1-b_1=a_2-b_2 + 2\pi k$, $\ell=1$ and $a_1+b_1=a_2+b_2 + 2\pi l$, where $k,l\in\mathbb Z$
 are such that \eqref{equ:timelike} is satisfied. Then for the initial conditions
 $\phi(x)=\sin(x)+\cos(x)$, $\psi(x)=-\cos(x)-\sin(x)$ there exists a unique
strengthened classical solution of problem \eqref{FIRST}--\eqref{LAST} and it is given by
 \[
  u(x,t)=u^D(x,t)= \sin(x-t) + \cos(x+t).
 \]
\end{example}

Indeed, conditions \eqref{equ:solvability} are satisfied because
$D_1(t)\equiv D_2(t) \equiv 0$. In addition it follows from
\eqref{equ:omega,nu} that $\omega(t)\equiv\nu(t)\equiv 0$.
In other words the solution is continuous and differentiable at the slits
(discontinuity steps are equal to zero).

Let us present another example when there exists a nontrivial
strengthened classical solution of problem
\eqref{FIRST}--\eqref{LAST} with nonzero discontinuity steps at the
slits.

\begin{example}
 Let $\ell=1$ and $h(t)\in C^\infty(\mathbb R)$ be such that
 $h(t)=0$ for $t\notin[0,1]$.
 For the initial conditions
 $\phi(x)=h(x+\alpha)$, $\psi(x)=-h'(x+\alpha)$, where $\alpha=b_2-a_2+1$,
 there exists a unique strengthened classical solution of problem \eqref{FIRST}--\eqref{LAST}
 and it is given by
 \[
  u(x,t)=u^D(x,t) + v_1(x,t) + v_2(x,t),
 \]
 where
 \[
 \begin{gathered}
  u^D(x,t)=h(x-t+b_2-a_2+1),\\
  v_1(x,t)=\theta(-b_1+a_1+t-x)h(1+b_1-a_1-t+x)\sign(x-a_1),\\
  v_2(x,t)=-\theta(-b_2+a_2+t-x)h(1+b_2-a_2-t+x)\sign(x-a_2).\\
 \end{gathered}
 \]
\end{example}
The solution is right-travelling wave, coming into the upper
slit and leaving out of the lower one.

To conclude
 we would like to note that one may interpret the
obtained conditions for initial data as saying that the classical
solution exists for ``almost all'' initial data from the functional
space of initial data. It would be interesting to study generalized
solutions of Cauchy problem on the Minkowski plane with the slits
and also to study the wave equation on more general non-globally
hyperbolic manifolds.

\section*{Acknowledgements}

This work was started at the seminar of Scientific Education Center
in the Steklov Mathematical Institute. It is partially supported by
Grants of RFBR 08-01-00727-a, NSh-3224.2008.1, AVCP 3341, DFG
Project 436 RUS 113/951. The work of O.V.G. is partially supported
by ``Leading Scientific Schools'' program NSh-691.2008.1.

 {\small

\end{document}